\documentclass[%
 reprint,
 amsmath,amssymb,
 aps,
]{revtex4-2}
\usepackage{graphicx}
\usepackage{dcolumn}
\usepackage{bm}
\usepackage{graphicx}
\usepackage{subcaption}
\usepackage{hyperref}
\usepackage{amsmath}
\usepackage{stackengine}
\usepackage{csquotes}
\usepackage{xcolor}
\usepackage{accents}
\usepackage[OT1,T2A]{fontenc}

\newcommand{\be}{\begin{equation}}
\newcommand{\ee}{\end{equation}}
\newcommand{\bea}{\begin{eqnarray}}
\newcommand{\eea}{\end{eqnarray}}

\makeatletter
\newcommand*{\rom}[1]{\expandafter\@slowromancap\romannumeral #1@}
\makeatother

\begin{document}


\title{On the derivations of the Equation of Hydrostatic Equilibrium    }
\author  {K\d{r}\d{s}\d{n}a  Dev}
\affiliation{%
 Department of Physics and Astronomy, Dickinson College, Carlisle PA
}%
\maketitle

\centerline{Abstract}
\noindent
We develop and present  a geometrical and an analytical  derivation of the  equation of hydrostatic equilibrium in spherically symmetric stars with a generalized stress tensor. The analytical derivation is based on the Navier-Cauchy equation. We also critically examine the  derivation of this equation found in textbooks on stellar astrophysics and show that there are errors in many of the derivations presented in textbooks. 


\section{Introduction}
On page 45 of his 1938  manuscript {\it An Introduction to the Theory of Stellar Structure} \cite{CHA}, Chandrasekhar wrote, \enquote{{\it Consider a perfect gas configuration in gravitational equilibrium. Then
\begin{equation*}
\frac{dp}{dr} = - \frac{G M(r)}{r^2} \rho   ~~~~~~~~~~~~~{\rm(55)}^{ \rm{\rom{1}}}
\end{equation*}
where $r$ denotes the radius vector with the center of configuration as origin. Furthermore, $M(r)$ is the mass inclosed inside a spherical surface of radius $r$ and $p = R \rho T$. The foregoing equation is an elementary consequence of hydrostatic equilibrium}}. A survey of textbooks  on stellar astrophysics \cite{CHA, PHIL, COU, KIPP,  HANS, DINA, CLAY, CANDO,  ZEIL,  RYD, MAOZ}, 
finds that the derivation of this  \enquote{elementary consequence of hydrostatic equilibrium}  is not an elementary task. We found that the derivations presented in textbooks including Chandrasekhar's book are incorrect and/or  logically inconsistent. 

The aim of this paper is to present logically consistent derivations of  the equation of hydrostatic equilibrium (EHE) in spherically symmetric Newtonian stars with a generalized stress tensor. We will develop both  an analytical and a geometrical derivation of the equation of the EHE
  using the principles of Newtonian mechanics and hydrostatic balance. The more widespread studies of the EHE considers  the pressure to be isotropic. In this paper we will take the pressure
to be anisotropic. A  spherically symmetric configuration   with anisotropic pressure is defined as a system where 
the radial pressure is not equal to the tangential pressure. An isotropic system is one in which the radial and tangential pressures are equal. 

  This paper is organized as follows. In the next section we will present our case that the derivations of the EHE based on spherical volume elements found in textbooks are wrong. In sections 3 we will describe our geometrical derivation. In section 4 we will return to the derivations found in textbooks and consider those  that use cylindrical volume elements.  In section 5 we will present an analytical derivation of the EHE based on the Navier-Cauchy equation. In section 6 we will summarize our work and give some general comments about derivations of the EHE found in textbooks.  
  
 In order to give a clear and detailed description of our reservations about various derivations  of the EHE found in textbooks  we will find it necessary to reproduce short  extracts from some of the books.  We will rewrite {\it verbatim} the extracts from the books and in order to avoid confusion between the  equations and figures  from the textbooks and our equations and figures and we will label the equations and figures from the textbooks in \textcolor{blue}{blue}. 
  
We will use $r, \theta$ and $\phi$ for spherical coordinates and $s, \phi$ and $z$ for cylindrical coordinates. 

\section{ Derivation of the EHE using a spherical volume element. }
All derivations of the equation of hydrostatic equilibrium in Newtonian stars relie on the principle of
  hydrostatic balance and some elementary geometrical considerations.  Hydrostatic balance is condition where there is equilibrium between the pressure forces and gravitational forces acting on a mass element in the fluid. There are two different geometrical derivations of the EHE that are found in textbooks. These derivations  differ from each other in the choice of the geometry of the volume element under consideration. The authors either study  the forces acting on a spherical volume element or a cylindrical volume element.  We will first consider  derivations  where spherical volume elements employed.
  
The derivations    of the EHE using a spherical volume element   found in  Phillips \cite{PHIL} and Choudhuri \cite{COU} are incorrect. Kippenhahn and Weigert \cite{KIPP} and  Hansen, Kawaler and  Trimble \cite{HANS} consider spherical shells in their derivations and their derivations also suffer from  {\it the curvature problem} as we will describe below. 

We will quote verbatim short extracts  from both Choudhuri's and Philips's books   and then we will discuss our reservations about their derivations.

 On page 62 of his book Choudhuri writes: {\it  Let us now consider a small portion of the shell between $r$ and $r + dr$. \textcolor{red}{If $dA$
is the transverse area of this small element, the forces exerted by pressure acting
on its inward and outward surfaces are $P dA$ and $-(P + dP) dA$, where $P$and
$P + dP$ are respectively the pressures at radii $r $and $r + dr$. So the net force
arising out of pressure is $-dP dA$, which should be balanced by gravity under
equilibrium conditions.} The gravitational field at r is caused by the mass M(r)
inside  $r$ and is equal to $-GM_r /r^2$. Since the mass of the small element under
consideration is $\rho dr dA$, the force balance condition for it is
\begin{equation*}
\textcolor{blue}{-dP dA - \frac{GM_r}{r^2} \rho dr dA = 0}
\end{equation*}
from which
\begin{equation*}
\textcolor{blue} { \frac{dP}{dr} =  - \frac{GM_r}{r^2} \rho  ~~~~~~~(3.2)}
\end{equation*}
This is the second of the stellar structure equations.} This is the end of the extract from Choudhuri's book. 

Phillips in his book on page 6 writes: {\it
We begin by finding an expression for the acceleration  of a mass element located at a distance  $r$ from the centre.  The matter  enclosed  by a spherical  shell  of  radius
$r$ has mass
\begin{equation*}
\textcolor{blue}{m(r) = \int_0^r \rho(r^{\prime}) 4 \pi (r^{\prime})^2 d r^{\prime}}
\end{equation*}
and  acts  as  a  gravitational  mass  situated  at  the  centre  giving  rise  to  an  inward gravitational  acceleration  equal  to
\begin{equation*}
\textcolor{blue}{g(r) = \frac{G m(r)}{r^2}}
\end{equation*}
\textcolor{red}{There is also, in general, a force arising form the pressure gradient. To find this we consider a small volume element located between radii $r$ and $r + dr$ of cross
sectional  area  $\Delta A$ and volume $\Delta r \Delta A$  as illustrated  in Fig 1.1  A net force arises if the pressure on the outer surface of the volume is not equal to the pressure on the
inner surface.  Indeed, the inward force on the volume  element  due to the pressure gradient  is}
\begin{eqnarray*}
 \textcolor{blue}{\left[ P(r)  + \frac{dp}{dr} \Delta r - P(r) \right] ~\Delta A 
 =  \left[ \frac{dP}{dr} \right] \Delta r \Delta A}
\end{eqnarray*}
Bearing in mind that the mass of the volume element is $\textcolor{blue}{\Delta M = \rho \Delta r \Delta A}$ we deduce that the inward acceleration of any element of mass at a distance r from the center of gravity and pressure is 
\begin{eqnarray*}
 \textcolor{blue}{-\frac{d^2 r}{d t^2}  = g(r)   + \frac{1}{\rho(r)} \frac{dP(r)}{dr}}
\end{eqnarray*}
Note that  to oppose gravity  the pressure  must  increase  towards  the centre.} This ends the extract from Phillips's book. 

\begin{figure}[h]
\centering
  \includegraphics[width=0.8\linewidth]{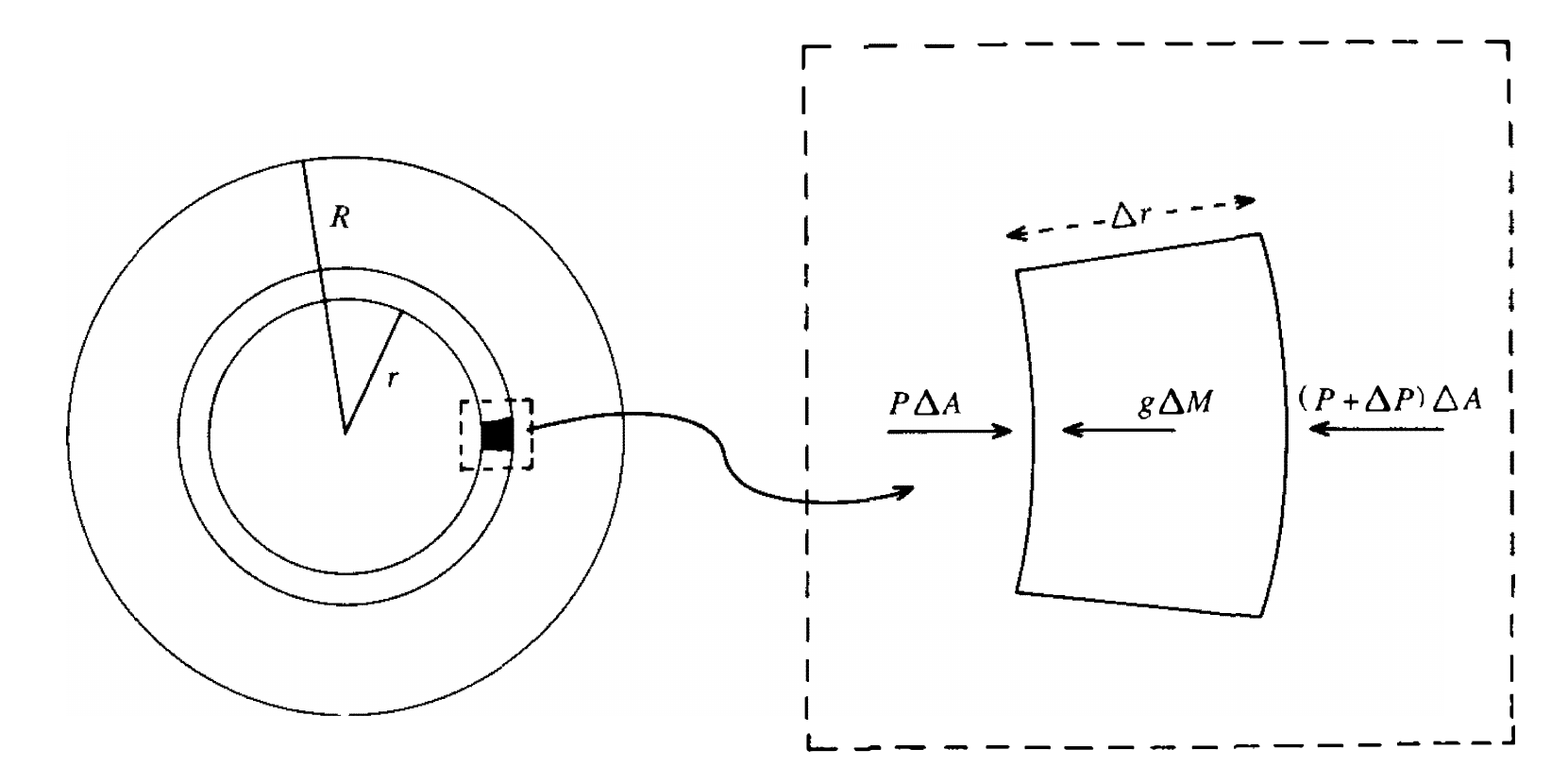}
  \caption{ \textcolor{blue}{ This figure illustrates the derivation of the EHE found in \cite{PHIL}. It is taken directly from page 6 the book. }}
   \label{phil}
\end{figure}
The derivations of the EHE presented  by  Phillips and Choudhuri  are wrong on two identical issues. We will refer to these as  (i) {{\it the curvature problem}} and (ii) {{\it the tangential force problem}}. 

{\it The Curvature Problem.}  The authors  considers the forces acting on spherical volume elements.  A figure from Phillips's book \cite{PHIL} is shown in fig \ref{phil}. They both claim that the force due to the radial pressure difference between the inner and outer surface area of the mass element is $dF(P_r)  = -dP_r dA$. (We have highlighted the relevant text  from the books  in {\textcolor{red}{red}.) This claim  is incorrect,
because  it  implicitly assumes that the areas of the two hemispherical caps
at $r$ and $r +   dr$ are  the same.  We note that if we take  the area at $r$ to be $dA$ as proposed in the textbooks  then the  area at $r + dr$ must be of the form 
be $ dA + \delta A$ in order to account for the difference in curvature of the two hemispherical caps and for $dF_{P_r}$ we should write
\begin{eqnarray}
\label{E1}
dF_{P_r} &= &   P_r(r + dr)  (dA + \delta A) - P_r(r)  dA  \\ \nonumber
 &=&   \left[ \frac{dP_r}{dr}  \right] dr dA + P_r(r)  \delta A.
\end{eqnarray}
Both terms on the right hand side  of the above equation are of the same order of smallness, thus
$ {\it a \, priori}$  there is no reason to neglect the term $P_r(r) \delta A$. It is standard practice however, in most  textbooks to take the area of the two caps to be equal. This is clearly wrong. We will refer to  this issue as the {{\it the curvature problem}}

{\it The Tangential Force Problem}.   {\it The tangential force problem} is the assumption is that the forces on the lateral  side of the cone sum to zero. This is not shown, it is considered to be obvious. However, we will show that
there is a nonzero force $F_t$, that arises from the pressure acting on the lateral side of the cone. This force $F_t$,  has a radial component that opposes the gravitational force on the volume element. In an extraordinarily fortuitous scenario  in fluids with  isotropic pressure the radial component of the force on the lateral side
cancel the $p(r) \delta A$ term that is missing  in (\ref{E1}). Thus the two errors, {{\it the area problem}}  and  {{\it the tangential force problem} }  compensate each other and the correct EHE is found for spheres with isotropic pressure. 

In the next section we will present a proper logical derivation of the EHE using a spherical volume element and then we will examine derivations based on a cylindrical volume element. We will show that the use of a cylindrical volume element in the derivation of the EHE is a clever attempt to solve the two problems the plague derivations based on spherical volume elements.

\section{Geometrical derivation of the equation for  hydrostatic equilibrium with anisotropic pressure.}

The equation of hydrostatic equilibrium  for the case with anisotropic pressure is
\be
\label{heq}
\frac{dP}{dr} =  - \frac{G\rho(r) M(r)}{r^{2}} + \frac{2}{r}(P_t - P_r)
\ee
\noindent here  $G$ is  Newton's gravitational constant.  The standard approach to the derivation of this equation is apply the principle of hydrostatic balance i.e., to equate
the pressure forces with gravitational force acting on a mass element.
\noindent We consider a mass element in a spherically symmetric configuration with anisotropic pressure.
In equilibrium the forces acting on a mass element $dm$ must sum to zero:
\be
{dF}_{g} + {dF}_{P_{r}} + {dF}_{P_{t}} = 0 ~,
\ee
\noindent where   ${dF}_{g}$ is the force due to gravity,  ${dF}_{P_{r}}$ is the force due to the radial pressure,  ${dF}_{P_{t}}$ is the force due to the tangential
pressure.
\begin{figure}
\centering
  \includegraphics[width=0.6\linewidth]{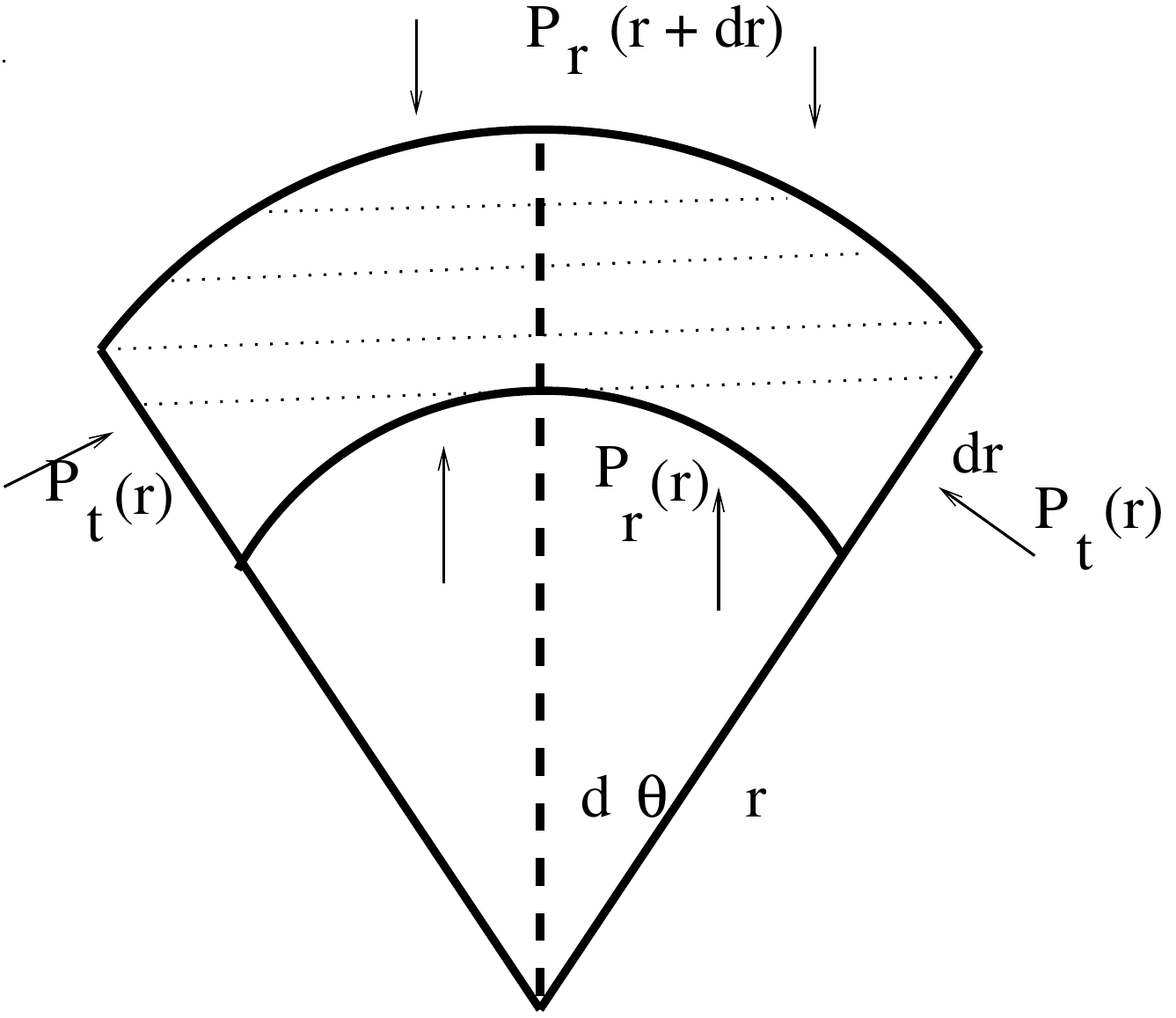}
  \caption{  A volume element $dv$ subjected to stresses from $P_{t}$ and $P_{r}$ }
   \label{tovad}
\end{figure}

\noindent  We will now derive expressions for  the forces. The gravitational force on a mass element $dm$ of volume  $dv$  located at a distance $r$ from the center of the star is
\be
dF_{g} = - \frac{G M \left( r \right) {\rm d}m}{r^2},
\ee
\noindent here
\be
\label{massdef}
M \left(r \right) = \int_{0}^{r}  4 \pi \rho \left(r \right) r^{2} dr
\ee
\noindent is the total mass contained in a sphere of radius $r$  and
\be
 dm = \rho(r) dv
\ee
\noindent is the mass of the fluid  element under consideration. Here we will choose the volume element to be a conical frustum that lies between $r$ and $r + dr$ with hemispherical caps.  The volume of a conical  frustum $dv_{cf}$ is obtained from the standard spherical volume  element $dv = r^2 \sin \theta dr d \theta d \phi$,  by integration over   $\phi$ i.e.,
 $dv_{cf} = 2 \pi r^2 \sin \theta dr d \theta$. Also since here, $\theta = d \theta$, then  $\sin \theta = d \theta$ and $dv_{cf} = 2 \pi r^2 dr (d \theta)^2$. The area of the hemispherical cap ar $r$ is $2 \pi r^2 \sin \theta d\theta  \approx 2 \pi r^2 (d \theta)^2$. A conical frustum is shown in fig. \ref{cone1}. This choice of volume element will allow us to readily compute the forces from the radial and tangential pressures. 

 Thus the gravitational force on the mass element is
\be
\label{fg}
dF_{g} = -  \rho  \frac{G  M(r) }{r^2}   2 \pi r^2  dr (d \theta)^2.
\ee
\noindent The force on the mass element due to the radial pressure is
\bea
\label{fpr}
dF_{P_{r}} &=& P_{r} ( r + dr)  dA(r + dr)
 - P_{r}   dA(r) \\ \nonumber
&=& (P_{r} + \frac{dP_{r}}{dr} dr) 2 \pi (r + dr)^{2}  (d \theta)^2  -  P_{r} 2 \pi r^2  (d \theta)^2 \\ \nonumber
&=&
 4 \pi  r   dr  (d \theta)^2  P_{r}  +  2 \pi r^2    dr (d \theta)^2  \frac{dP_{r}}{dr}.
  \eea
\noindent If  we consider only the forces due to gravity and the radial pressure on the mass element then we find that the equation for hydrostatic equilibrium is
\be
\begin{split}
  4 \pi  r  dr  (d\theta)^2 P_{r}  +  2 \pi r^2 dr (d\theta)^2  \frac{dP_{r}}{dr}  = ~~~~~~~~~~~~~~ \\
  -  \rho  \frac{G  M(r) }{r^2}   2 \pi r^2  dr (d\theta)^2.
  \end{split}
\ee
\noindent This equation simplifies to
\be
\label{aa}
\frac { dP_{r}}{dr}  + \frac{2}{r} P_{r}   = -   \frac{\rho(r)G  M(r)}{r^2} .
\ee
If the second term on the left hand side in eq. (\ref{aa}) was not present  then resulting equation is standard equation for hydrostatic equilibrium
in stars with isotropic pressure. This second term is present because	 the  the area of hemispherical caps at $r$ and $ r + dr$ are not equal. This
is a crucial point: {\it because the curvature of the the two hemispherical caps are different, their areas are also different}. In several
textbooks on stellar astrophysics this point is not noted  and the authors derive  the  equation for hydrostatic equilibrium
by making two erroneous assumptions: (i)  the areas of the two hemispherical caps are equal, ({{\it the curvature problem}}) and (ii) the contribution to the total force from the tangential pressure acting on the lateral side of the cone
is zero ({{\it the tangential force problem}}). We have shown above that first assumption is incorrect and below we will demonstrate that there is a non-zero contribution to the net force from the tangential pressure acting on the lateral side of the of the cone. This extra term exactly cancels the second term on the
left hand side of   (\ref{aa}), when the pressure is isotropic. If  the pressure is anisotropic the two terms do not cancel and we arrive at the equation for hydrostatic equilibrium with anisotropic pressure  (\ref{heq}).

\begin{figure}
\centering
  \includegraphics[width=0.5\linewidth]{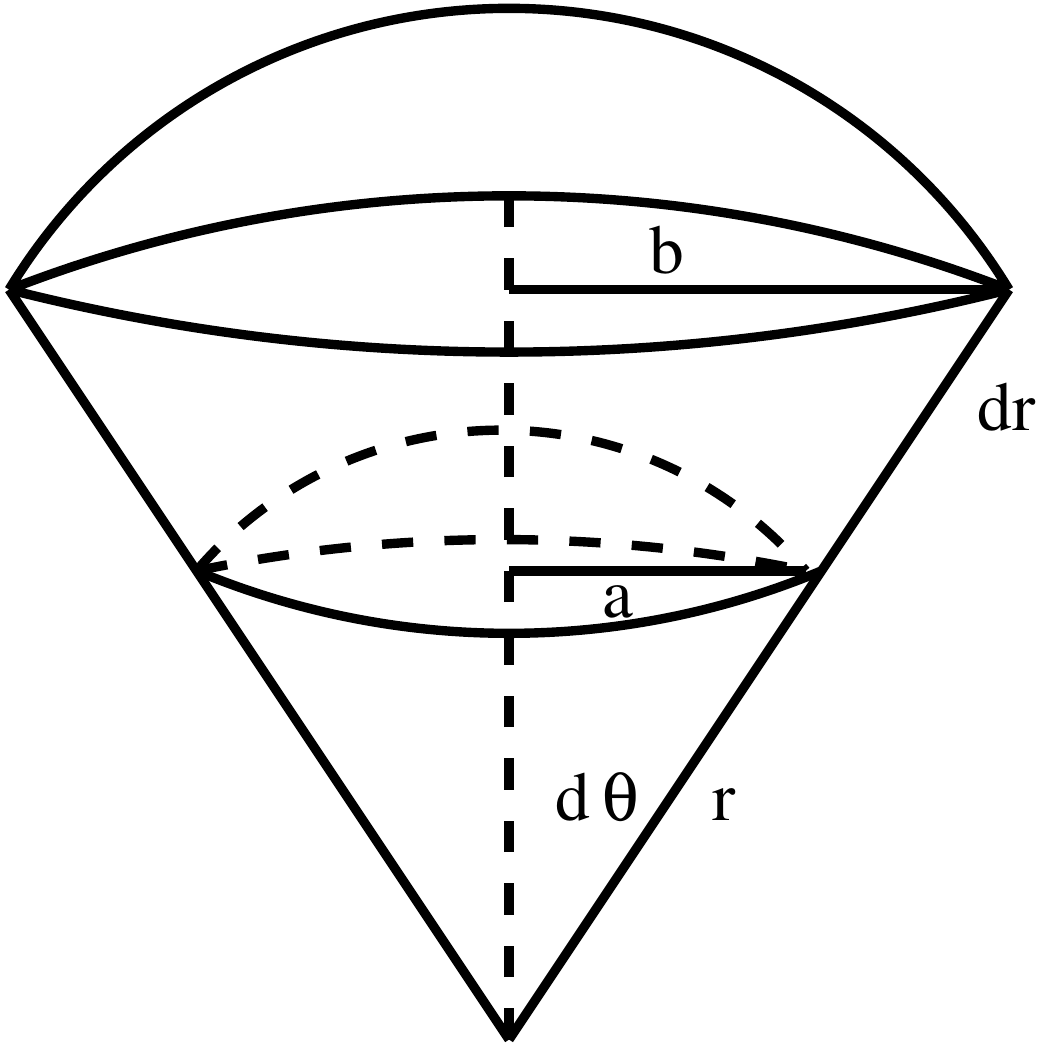}
 \caption {A conical frustum located between $r$
and $r + dr$}
    \label{cone1}
\end{figure}

\begin{figure}
\centering
  \includegraphics[width=0.7\linewidth]{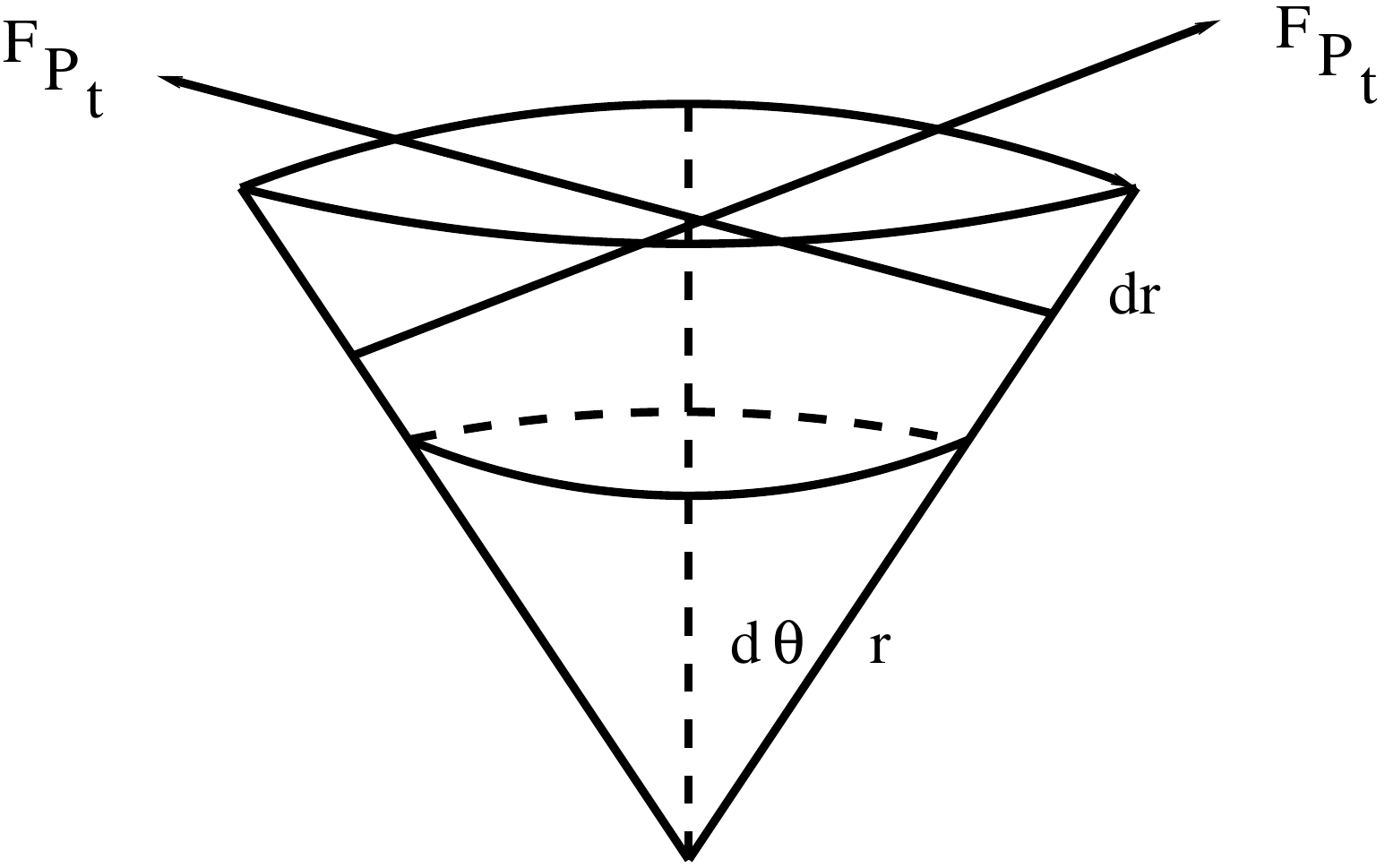}
 \caption{The force due on conical frustum located between $r$
and $r + dr$ due to $P_{t}$}
    \label{coneb}
\end{figure}

 \noindent The force due to the tangential pressures $P_{\theta}$ and $P_{\phi}$ is
\begin{eqnarray}
\label{fpt1}
dF_{P_t} =  (P_{\theta} + P_{\phi} )  dA_{cf}
\end{eqnarray}
\noindent where $dA_{cf}$ is the  lateral area of a conical frustum between $ r$ and $ r + dr$. From fig \ref{cone1} we find that
\be
\begin{split}
dA_{cf}  = \pi  ( a + b) \times {\rm length\, of \,slant \,side}  ~~~~~~~~~\\
~~~~~= \pi ( r d\theta + (r + dr) d\theta) dr = 2 \pi  r d\theta  dr
\end{split}
\ee 
 Since $P_{\theta} = P_{\phi} = P_t$  for any spherical symmetric configuration, the component of $dF_{P_t} $ that acts along the radial direction is  
 \begin{eqnarray}
\label{fpt}
(dF_{P_t})_r =  4  \pi r  dr (d \theta)^2 P_{t}   ~~~~~
\end{eqnarray} 
 Hydrostatic equilibrium requires
\be
dF_{P_r}  +  (dF_{P_{t}})_r =  dF_g.
\ee
Substituting for $dF_{P_r}$ (\ref{fpr}),    $ (dF_{P_{t} })_r$ (\ref{fpt}), and  $  dF_g$ (\ref{fg}), we find that 
\be
\begin{split}
 4 \pi  r  dr  (d\theta)^2 P_{r}  + 2 \pi r^2 dr (d\theta)^2  \frac{dP_{r}}{dr} ~~~~~~~~~~~~~~~~~~~~ \\  -  4  \pi r  dr (d \theta)^2 P_{t}                  =  
  -  \rho  \frac{G  M(r) }{r^2}   2 \pi r^2  dr (d\theta)^2.
  \end{split} 
\end{equation}
Simplifying the  above expression we find:
\bea
\label{new}
\frac{dP_{r}}{dr} =  - \frac{\rho(r)G  M(r)}{r^{2}}
+ \frac{2}{r}(P_{t} - P_{r}).
\eea
\noindent This is the equation of hydrostatic equilibrium for Newtonian stars  with anisotropic pressure. In the next section we return to our critique of the derivations of the EHE found in textbooks and consider derivations based on cylindrical volume elements.

\section{ The derivation of the EHE using a cylindrical  volume element.}
 Cylindrical volume elements are mentioned in the derivations found in the following textbooks: Chandrasekhar \cite{CHA}, Prialnik \cite{DINA}, Clayton \cite {CLAY}, Carroll and Ostlie \cite{CANDO}, Zelik, Geogory and Smith \cite {ZEIL}, Ryden and Peterson \cite{RYD} and Maoz \cite{MAOZ}. 
 
 We will first reproduce extracts from Clayton's and Maoz's books and then we describe our basic reservations about this and other  derivations based on cylindrical volume elements. 

\begin{figure}[h]
\centering
 \includegraphics[width=0.6\linewidth]{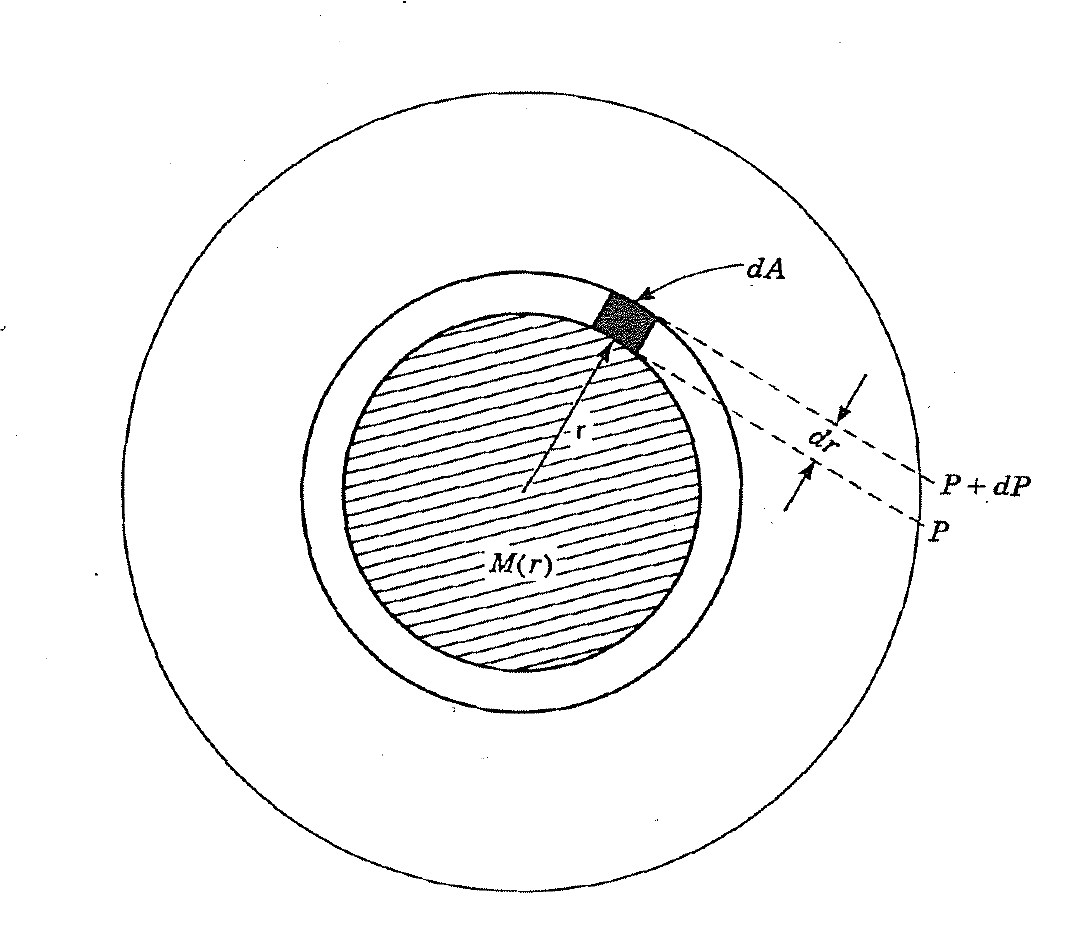}
   \caption{ \textcolor{blue}{This figure is used in the derivation of the EHE found in Clayton's book \cite{CLAY}. It is labelled  \textcolor{blue}{Fig 2.14} in the book.}}
   \label{nut}
\end{figure} 

Clayton writes on  page 130  of his book:  \enquote{{\it Consider the small cylindrical volume element with axis of length $dr$ parallel to the radius vector at the point $r$ and having cross-sectional area equal $dA$. Let $dP$ be the pressure increment associated with $dr$ Then the radial force on this volume element due to the pressure differential is
\begin{equation*}
\textcolor{blue}{F_P = Pda - (P + dP)dA = - dP dA  ~~~~~~(\operatorname{2-177})}
\end{equation*}
since the pressure will actually decrease in the radial direction, the differential $dP$ will be negative and the pressure force will then be positive. If the volume element is not to be accelerated upward by $F_P$, it is necessary that it be exactly cancelled by the central gravitational force in the volume element.  If $dm$ is the mass of the volume element and $M(r)$ is the mass interior to the spherical surface at radius $r$, the gravitational force is 
\begin{equation*}
\textcolor{blue}{F_g  = - G \frac{M(r) dm}{r^2}   ~~~~~~~~~~~~~~(\operatorname{2-178})          }
\end{equation*}
where 
\begin{equation*}
\textcolor{blue}{M(r) = \int_0^r \rho(r) 4 \pi r^2 dr  ~~~~~~~~~~    (\operatorname{2-179})             }
\end{equation*} 
The gravitational constant $G = 6.670 \times10^{-8} \, dyne\,cm^2\,g ^{-2}$. In writing these equations, we have clearly considered only the spherically symmetric problem. The important problem of rotation is discussed in Chap. 6. Because the mass of the small volume element is $\rho dA  dr = dm$, we have 
\begin{equation*}
\textcolor{blue}{ 0  = F_P + F_G = -\frac{dP}{dr} - G \frac{\rho M}{r^2} ~~~~~    (\operatorname{2-180})                     }
 \end{equation*}
 This force balance together with the definition of ${\rm\textcolor{blue}{ Eq. (\operatorname{2-180})}}$, constitutes one of the most important condition to be satisfied by stellar structures.}}

This ends the extract from Clayton's book. A shorter but similar discussion is given in Maoz's book \cite{MAOZ}. Maoz writes on page 23 of his book:
\begin{figure}[h]
\centering
 \includegraphics[width=0.6\linewidth]{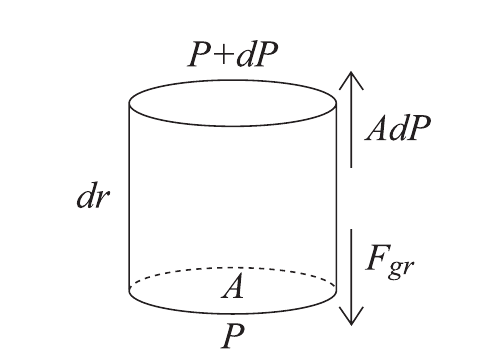}
   \caption{ \textcolor{blue}{This figure is used in the derivation of the EHE found in Maoz's book \cite{MAOZ}. It is labelled  \textcolor{blue}{Figure 3.1} in the book.}}
   \label{Maoz}
\end{figure}
{\it Consider now a small, cylinder-shaped mass element inside a star, with A the area of the cylinder's base,
and $dr$ its height see \textcolor{blue}{Fig. (3.1)}. If there is a pressure difference $dP$ between the top of the
cylinder and its bottom, this will lead to a net force $A  dP$ on the mass element, in addition
to the force of gravity toward the center. Equilibrium will exist if
 \begin{equation*}
\textcolor{blue}{ - \frac{G M(r) dm}{r^2}  - A dP = 0~~~~~~~~~~~~(3.17)} 
 \end{equation*}
 But
 \begin{equation*}
\textcolor{blue}{ dm = \rho(r) A dr, ~~~~~~~(3.18) }
\end{equation*}
which leads us to the first equation of stellar structure, the equation of hydrostatic equilibrium 
  \begin{equation*}
\textcolor{blue}{  \frac{dP(r)}{dr} = - \frac{G  \rho(r) M(r)}{r^2} ~~~~~~~~~~~~(3.19)} 
 \end{equation*} }
 \noindent This is the end of the extract from Maoz's book.

We find that derivations of the EHE that use cylindrical volume elements carefully circumnavigate  the problems encountered when using spherical volume elements, i.e.,  {\it the curvature problem} and {\it the tangental force problem}.  We can see this by examining  fig \ref{nut} taken  from Clayton's book and fig \ref{Maoz} taken from Maoz's book. 
 An inspection of the diagrams  shows that the   surface area of the caps of cylindrical volume elements are flat discs.
  Thus, this geometry immediately eliminates the {{\it the curvature problem}}. Further, since pressure  forces always act directly perpendicular to a given surface, the pressure forces on the curved surface of the cylinder will sum to zero and this solves the   {{\it the tangential force problem}}.     However, we note that in the interior of the stellar configurations that we are considering  both the pressure and the gravitational field have spherical symmetry. Thus, {\it  the use of volume elements with cylindrical symmetry in a spherical symmetric medium  should be problematic}. We have some reservations about the use of cylindrical volume elements in the derivation of the EHE. We will refer to these as   (i) {\it the area problem} and (ii) {\it  the anisotropy problem.} 
 

(i) {\it The  area problem.} We found in our studies of the various derivations of the EHE using cylindrical volume elements that all authors arrive at an expression of the following form:
\begin{equation*}
\textcolor{blue}{ - \frac{G M(r)}{r^2} dr A  - dPA = 0.} 
 \end{equation*}
 This is \textcolor{blue} {Equation 3.17}  in Maoz's book. The term \textcolor{blue}{$dP A$} represents the force that arises due to the pressure difference between the caps. {\it An implicit assumption here is that the radial pressure i.e.,   a spherically symmetric quantity is constant over flat surfaces with cylindrical symmetry}. Maoz does not explicitly state this in his derivation but he could not have written his expression for the pressure force without this assumption. Similarly, Clayton  could only arrive  at \textcolor{blue}{$F_P = - dP dA$} in his \textcolor{blue}{Eq (2-177)} using  this assumption. 
 However, we disagree with this assumption.  Our reasoning is as follows. We are asked to consider a cylindrical volume element  of height $r$ and surface area $dA$. The change in height $dr$ is necessary for there to exist a radial pressure gradient between the top and bottom of the cylinder,  i.e.,   when  $r \rightarrow r + dr$, then $P(r + dr)  -P(r)  =  (dP/dr) dr \neq 0$.  We do not object to this preposition.
 
 \begin{figure}[h]
\centering
 \includegraphics[width=0.8\linewidth]{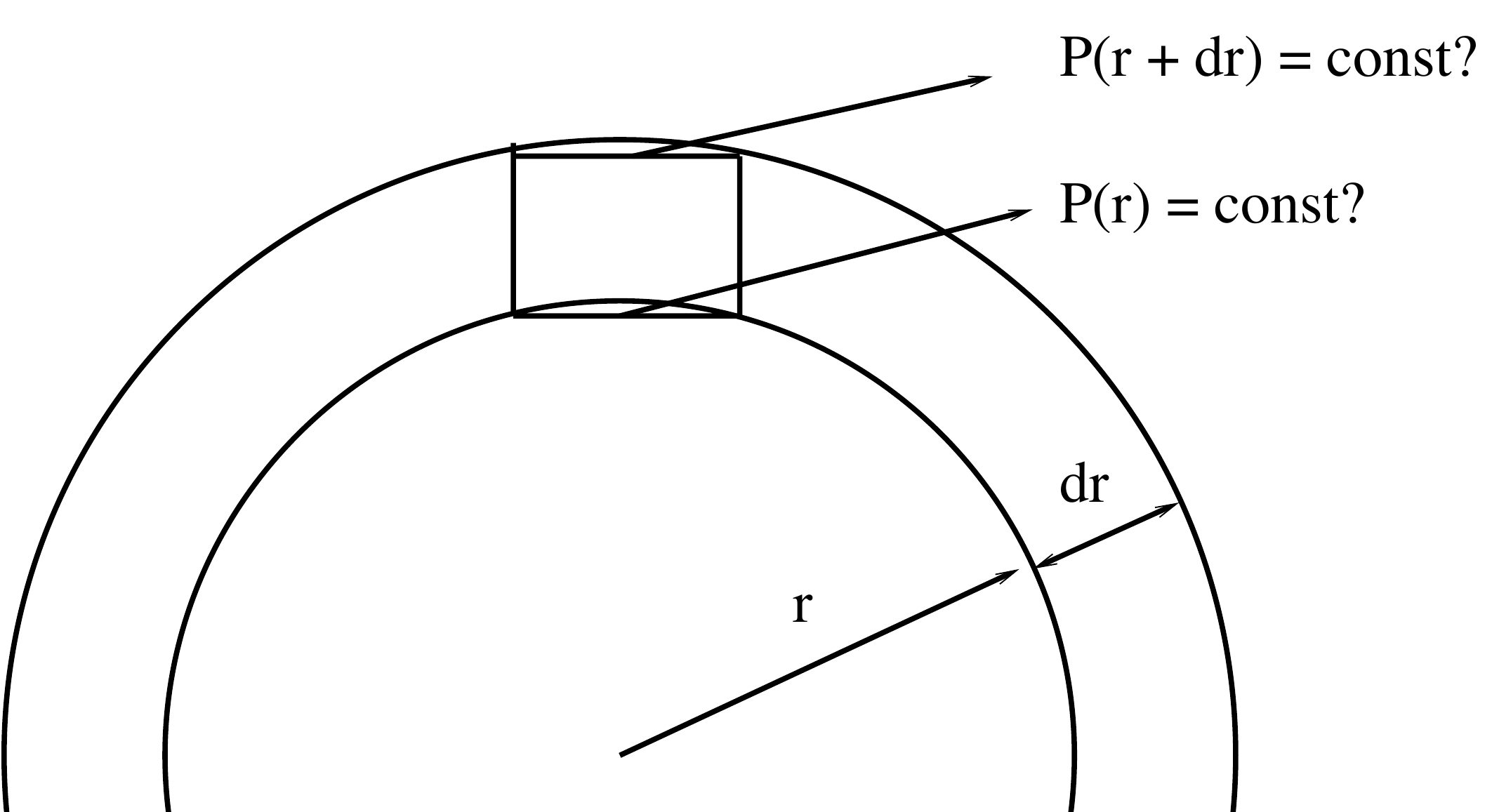}
   \caption{ This figure illustrates {\it the area problem}}
   \label{circle}
\end{figure}

 However, we have a conceptual difficulty in understanding how is it possible for the radial pressure, a spherically symmetric quantity, 
  to be constant over a flat area that is perpendicular to the radial direction (see fig \ref{circle}). The pertinent question here is, how small is the area $dA$?  Certainly, it must be at least $\pi (ds)^2$, this is because the smallest possible value the cylindrical radial coordinate $s$ can have is $ds$.  Any quantity  smaller than $ds$  is a point. Now we have established that  $dA$ must have a radius of minimum length  $ds$, we ask why is there a pressure difference with a change in height $dr$ but there is no pressure difference along the surface when $s$ changes by at least $ds$, certainly when $s \rightarrow s + ds$, we must have $P(r(s + ds)) - P(r(s)) =  (dP/dr)(dr/ds) ds \neq 0$. If $ds  = 0$ then we are not considering an infinitesimal area, the geometrical object under consideration is  a point and the derivation is moot.  We will call this  {\it the area problem}.

The relevant question now  in view of  {\it the area problem}, is what form does the EHE assume when using a cylindrical volume element? The expression for the gravitational force on the mass element is undisputed. We now need to find an expression for the force due to the pressure difference.

The gravitational force on the cylindrical mass element  $dm = \rho dv$ is 
\begin{eqnarray}
 F_g & = &   -  \frac{GM(r)}{r^2} dm   = -  \frac{GM(r)}{r^2}  \rho dv \\ \nonumber 
 &=&   -   \frac{G M(r) \rho}{r^2} dz A_c  \approx    -   \frac{ G M(r) \rho }{r^2} dr A_c. 
  \end{eqnarray}
 In the last line in the above equation we have used the approximation $dz =  dr$.
 \noindent The force due to the pressure on the bottom  cap is given by an expression of the form
 \begin{eqnarray}
  F_{P_b} &=& \int P(r(s)) dA_c   \\ \nonumber
  &=& \int \left( P(r)  + \frac{dP(r(s))}{ds} ds\right) dA_c.
\end{eqnarray}
Here $dA_c$ is an area element  of  the flat cap. The force due to the pressure on the top  cap is given by an expression of the form
 \begin{eqnarray}
  F_{P_t} &=& \int P(r(s) + dr(s)) dA_c \\ \nonumber
 &=&   \int \left( P(r)  +  \frac{dP(r(s))}{ds} ds  + \frac{dP(r)}{dr} dr  \right. \\ \nonumber 
 & & ~~~~~~\left. + \frac{d^2P(r(s))}{ds dr} ds dr \right) dA_c. 
 \end{eqnarray}
 The net force due to the pressure difference is 
 \begin{eqnarray}
 dF_P &=&  \int \left(     \frac{dP(r)}{dr} dr  +   \frac{d^2P(r(s))}{ds dr} ds dr\right) dA_c  \\ \nonumber
  &=& \frac{dP}{dr} dr A_c + \int \frac{d^2P(r(s))}{ds dr} ds dr dA_c.
  \end{eqnarray}  
For equilibrium we need 
\be
F_g + F_P =0.
\ee
Substituting into this equation we find 
\be
  -   \frac{ GM(r)  \rho}{r^2}  dr A_c -    \frac{dP}{dr} dr A_c  - \int \frac{d^2P(r(s))}{ds dr} ds dr dA_c = 0.
 \ee
This  the EHE  for a cylindrical volume element  in a spherically symmetric system.  We will obtain the standard EHE in a spherically symmetric medium if $dP/ds = 0$, however, if  this condition holds, then the caps have a surface area equal to zero and the cylinder is reduced to a needle.



 
  (ii). {\it The anisotropy problem.} We note that  despite our dedicated efforts, we were unable to  derive an EHE for spherically symmetric systems with anisotropic pressure using a cylindrical volume element. This is probably not surprising, since a cylindrical volume element has no surface over which the components of a spherically symmetric  pressure tensor are directly perpendicular or parallel. Thus the pressure forces on each surface is due to a mixture of both the radial and tangential pressures acting on those surfaces. We think it is very challenging mathematical problem to obtain the correct EHE for systems with spherical symmetry and anisotropic pressure by studying the forces on a cylindrical volume element. This  is {\it the anisotropy problem}.

 We consider the use of a cylindrical volume element in the derivation of the EHE to be a  \enquote{\it geometrical ruse} designed to circumnavigate  the problems encountered when using spherical volume elements, i.e.,  {\it the curvature problem} and {\it the tangental force problem}, and not  a serious logically consistent mathematical derivation of the EHE.

\section{An analytical  derivation of the equation for  hydrostatic equilibrium with anisotropic pressure.}

A pertinent question to ask is since the EHE is such a well known and extensively  studied equation, does there exist  proper analytical derivations of this equation that does not rely on geometrical subtleties and \enquote{\it ruses} ?  The answer is yes, such derivations do exist. Examples of analytical derivations can be found in Collins \cite{COLL} and Thorne and Blandford \cite{THOR}. 

 In his derivation  Collins  studies the flow of fluids. He  starts with the Boltzmann transport equation and develops the Euler-Lagrange equations  of hydrodynamic flow:
 \be
 \label{col} 
 \frac{\partial \vec{u}}{\partial t}  + (\vec{u} \cdot \vec{\nabla}) \vec{u}  = - \vec{\nabla} \Phi - \frac{ \vec{\nabla} \cdot{\overleftrightarrow{P}}} {\rho}
 \ee
 Here $\vec{u}$ is the flow velocity of a particle, $\Phi$ is the potential associated with an external central force acting on the particles and $\overleftrightarrow{P} $ is the pressure tensor. When  the flow velocity is constant  and isotropic the entire left-hand side of  (\ref{col})  is zero and we obtain 
 \be
  \vec{\nabla}\cdot{\overleftrightarrow{P}} =  -  \rho \vec{\nabla} \Phi 
 \ee 
 This is the EHE and here it is an expression for  the conservation of momentum. 

Thorne and Blandford in their approach to  derivation of the EHE, applies the ideas and concepts developed for studying elastostatics in solids  to the study of hydrostatics in fluids. The fundamental equation of elastodynamics is the Navier-Cauchy equation. The Navier-Cauchy equation describes the displacements in an elastic solid under the influence of external forces or internal stresses. The Navier-Cauchy equation can be written as 
\be
\overrightarrow{\nabla}\cdot  { { {\overleftrightarrow{\sigma}}} }+ \vec{f}  = \rho \frac{ d^2 \vec{\xi}}{d t^2}.
\ee
Here ${ \overleftrightarrow{\sigma}}$ is the Cauchy stress tensor, $\vec{f}$ is external force density and $\vec{\xi}$ is the displacement vector. Elastostatic equilibrium is achieved when the time derivatives  are set to zero. Thus the fundamental equation of elastostatics is
\be
\label {E1a}
\overrightarrow{\nabla}\cdot  { {{\overleftrightarrow{\sigma}}} }      = - \vec{f}
\ee
This equation expresses elastostatic balance between the internal stresses in the solid and the external applied force(s). In hydrostatics a similar balance is achieved between the pressure forces and the applied external forces. The fundamental equation of hydrostatics has the same form as (\ref{E1a}) except now the Cauchy stress tensor is replaced by the fluid's stress tensor, $\overleftrightarrow{T}$. Thus the fundamental equation of hydrostatics is 
\be
\label {E1c}
\overrightarrow{\nabla}\cdot   { {{\overleftrightarrow{T}}} }         = - \vec{f}
\ee
\noindent 
We are considering  fluids with anisotropic pressure that have  no shear or vorticity thus here the stress tensor has a diagonal form :
\be
 {\overleftrightarrow{T}} = \begin{bmatrix}
 				 P_r(r) & 0 & 0 \\
				 0 & P_t(r)  &0 \\
				 0 & 0 & P_t (r) 
				 \end{bmatrix}
\ee

In the interior of stars the only body force acting  is gravity and it has only a radial component. In evaluating	(\ref{E1c}) we find that only  the $r$ component of 
$  \overrightarrow{\nabla}\cdot  { { {\overleftrightarrow{T}}} }   $ gives a non-trivial expression. The $\theta$ and $\phi$ equations gives zero on both sides of the equations.   Writing out the $r$ component of (\ref{E1c}) we find 
\be
\begin{split}
\label{E1d}
\frac{\partial T_{rr}}{\partial r} + \frac{1}{r} \frac{\partial T_{r\theta}}{\partial \theta} + \frac{1}{r \sin \theta} \frac{\partial T_{r\phi}}{\partial \phi} ~~~~~~~~~~~~~~\\
 + \frac{1}{r} ( 2 T_{rr} - T_{\theta \theta} - T_{\phi \phi} + T_{r \theta} \cot \theta) = - f.
\end{split}
\ee
 Substituting  $f_r = \rho g$  ($g$ is the gravitational acceleration), and taking into consideration the form of  $ {\overleftrightarrow{T}} $ we find that (\ref{E1d}) reduces to 
\be
\label{E1b}
\frac{\partial P_r(r)}{\partial r}  + \frac{1}{r} ( 2 P_r(r)  - 2 P_t(r)) = - \rho g
\ee
 In the interior of a spherically symmetric star at a distance $r$ from the center 
 \be
 g =  \frac{G M (r)}{r^2}
 \ee
 Substituting this expression for $g$ in (\ref{E1b}) and rearranging terms  we find
\bea
\label{new}
\frac{dP_{r}}{dr} =  - \frac{\rho(r)G  M(r)}{r^{2}} + \frac{2}{r}(P_{t} - P_{r})
\eea
\noindent This is the equation of hydrostatic equilibrium for Newtonian stars  with anisotropic pressure. Thus we have derived the EHE analytically without the need to consider geometrical subtleties.

\section{Conclusion}
We have given two derivations of the equation of hydrostatic equilibrium. The first derivation  applied the  principles of Newtonian mechanics and simple geometrical arguments to obtain the EHE. 

Our second derivation is purely analytical and it based on applying the principles of elastostatics in solids to hydrostatics in fluids. We have also shown that the derivation of this equation as it is found in almost all textbooks on stellar astrophysics is flawed.

Finally, our advice for authors who find it necessary to include a derivation of the EHE in their manuscripts is to either adopt the correct derivations presented in this paper and elsewhere \cite{THOR}, or follow the path taken by Bhom-Vitense \cite{BHOM} or Winn \cite{WINN}. They both present an elementary derivation of the equation of  hydrostatic equilibrium in a constant gravitational field that is found in introductory physics textbooks and then generalize the EHE to situations with spherical symmetry. The EHE is well known, there is no need to present incorrect derivations or resort to \enquote{\it geometrical ruses}.

\section{Acknowledgements}
We thank Lars English for reading and editing the manuscript.

\end{document}